\documentclass[notoc,paper,letterpaper]{JHEP3}
\usepackage{epsfig,amssymb} 
\newcommand{\be}{\begin{equation}}
\newcommand{\ee}{\end{equation}}
\newcommand{\bea}{\begin{eqnarray}}
\newcommand{\eea}{\end{eqnarray}}
\newcommand{\ie}{{\it i.e.}}
\newcommand{\eg}{{\it e.g.}}

\newcommand{\Z}{\mathbb{Z}}
\newcommand{\U}{\mathop{\rm U}}
\newcommand{\tr}{\mathop{\rm tr}}
\newcommand{\SU}{\mathop{\rm SU}}
\newcommand{\SO}{\mathop{\rm SO}}

\newcommand{\Sp}{\mathop{\rm Sp}}
\newcommand{\id}{\hbox{1\kern-.23em{\rm l}}}
\def\Pf{\mbox{Pf}\,}
\def\det{\mbox{det}\,}
\def\bar{\overline}
\def\t{\tilde}
\def\h{\hat}

\def\wh{\widehat}
\def\del{{\partial}}

\def\MM{{\cal M}}
\def\NN{{\cal N}}
\def\SS{{\cal S}}
\def\a{\alpha}
\def\b{\beta}
\def\g{\gamma}
\def\d{\delta}
\def\e{\epsilon}
\def\ve{\varepsilon}
\def\z{\zeta}
\def\th{\theta}
\def\k{\kappa}
\def\l{\lambda}
\def\L{\Lambda}
\def\m{\mu}

\def\s{\sigma}

\def\ad{{\dot\a}}

\def\veb{{\bar\ve}}
\def\thb{{\bar\th}}
\def\lb{{\bar\l}}
\def\mb{{\bar\m}}

\def\nf{{n_f}}
\def\nc{{n_c}}
\def\we{{W_{\rm eff}}}
\def\wep{{W^\ve_{\rm eff}}}

\def\Db{{\bar D}}
\def\dM{{\d M}}
\def\dMb{{\d\Mb}}
\def\Mb{{\bar M}}
\def\Mo{{M_0}}
\def\Mbo{{\Mb_0}}
\def\Mvo{{M^\ve_0}}

\title{On singular effective superpotentials in supersymmetric
gauge theories}

\author{Philip C. Argyres and Mohammad Edalati\\
Department of Physics, University of Cincinnati, 
Cincinnati OH 45221-0011\\
\email{argyres,edalati@physics.uc.edu}}
\abstract{We study $N=1$ supersymmetric $\SU(2)$ gauge theory in 
four dimensions with a large number of massless quarks.  We argue 
that effective superpotentials as a function of local gauge-invariant 
chiral fields should exist for these theories.  We show that 
although the superpotentials are singular, they nevertheless correctly 
describe the moduli space of vacua, are consistent under RG flow to 
fewer flavors upon turning on masses, and also reproduce by a 
tree-level calculation the higher-derivative F-terms calculated by 
Beasely and Witten \cite{bw0409} using instanton methods.  We note 
that this phenomenon can also occur in supersymmetric gauge theories 
in various dimensions.}

\begin{document}

\section{Introduction and summary}

Using the selection rules of four dimensional $N=1$ supersymmetry, 
exact results for superpotentials for supersymmetric gauge theories 
have been obtained; see \cite{is9509,p9702} for short reviews.  
These results have been inferred in field theory by an elaborate 
series of consistency checks, having largely to do with consistency 
upon integrating out massive chiral multiplets.  The basic strategy 
for finding these results has been a loose kind of induction in the 
number of light flavors in which one works one's way up to larger 
numbers of light flavors by making consistent guesses.  The heuristic 
picture obtained in this way for superQCD is that quantum effects are 
more pronounced in the low energy effective action the fewer the number 
of light flavors. 

It is natural to ask whether this procedure can be made more deductive 
and uniform by turning it on its head, and starting instead with the IR 
free theories with many massless flavors.  Since the leading low energy 
effective action of IR free theories are free, how do they manage to 
generate the strong quantum effects as flavors are integrated out?  
Recently, C. Beasely and E. Witten \cite{bw0409} have shed light on how 
quantum effects at low number of flavors are inherited from the 
large-flavor theory.  They found that at large number of flavors there 
are higher-derivative $F$-terms (of a special form) in the action which, 
upon integrating out, descend to lower-derivative operators, until they 
finally become relevant, and manifest themselves as quantum corrections 
in the low energy effective action (\eg\ as a quantum deformation of the 
moduli space).  

They compute these terms in $\SU(2)$ superQCD with an 
arbitrary number $\nf$ of fundamental flavors by a one-instanton 
argument.  This is done intrinsically on the moduli space, \ie\ using 
only the massless multiplets in the vicinity of an arbitrary non-singular 
point on the moduli space.  But, interestingly, for $\nf=2$ and $3$ they 
also show that the higher-derivative terms can be derived by simply 
integrating out massive modes at tree-level from an effective 
superpotential defined on a larger configuration space made up of vevs 
of the local gauge-invariant chiral meson field.

This raises the question of whether a similar efficient description
of larger-flavor cases can be made in terms of effective superpotentials.
Now, such superpotentials are thought to be problematic because, for 
large enough number of flavors, they are singular \cite{s9402,ip9505} 
when expressed in terms of local gauge-invariant chiral vevs, even 
away from the origin.  Also, these superpotentials do not vanish as 
the strong-coupling scale of the theory $\L$ vanishes.  Indeed, such 
an effective superpotential need not even exist \cite{s9402};  for only 
if there is a region in the configuration space of the chosen chiral 
vevs where all of them are light together and comprise all the light 
degrees of freedom, are we then assured that there is a Wilsonian 
effective action in terms of these fields in that region.
If this condition is satisfied, then the resulting effective 
superpotential can be extended over the whole configuration space by 
analytic continuation using the holomorphy of the superpotential. 
For large-flavor superQCD, the only region where all the components 
of the meson and baryon fields become light at the same time is at the 
origin.  But 't Hooft anomaly matching conditions imply 
\cite{s9411} that there 
must be additional extra light degrees of freedom beyond the meson and 
baryon fields at the origin.  Thus no superpotential written 
solely in terms of mesons and baryons need exist, for there
is no guarantee that modes of other operators which account for the
additional massless degrees of freedom at the origin are not as light
as the meson and baryon modes, and so must also be included in a
consistent effective action. 

However, when there are so many massless flavors that the theory is 
IR free, we know what the light degrees of freedom are near the origin, 
since we have a weakly coupled lagrangian description there.  The 
physics can be made arbitrarily weakly coupled simply by taking all 
scalar field vevs $\m \sim \langle\phi\rangle << \L$ where $\L$ is the 
strong coupling scale (or UV cutoff) of the IR free theory.  In this 
limit the physics is just the classical Higgs mechanism, and all 
particles get masses of order $\m$ or less.  The Wilsonian effective 
description results from integrating out modes with energies greater 
than a cutoff, which we take to be some multiple of $\m$.  The effective 
action will then include all local gauge-invariant operators made 
from the fundamental fields in the lagrangian and which can create
particle states with masses below the cutoff.  For the purpose of 
constructing the effective superpotential, the relevant local 
gauge-invariant operators are those in the chiral ring. 
It is then just a matter of constructing a set of operators in 
the classical gauge theory which generates the chiral ring.  An 
effective superpotential which is a function of these operators must 
then exist.

To be concrete, consider the simplest example, which will be the focus
of this paper: $N=1$ supersymmetric $\SU(2)$ QCD in four dimensions.
This theory has an adjoint vector ``gluon" multiplet $W^{(ab)}_\a$ and 
$2\nf$ fundamental ``quark" chiral multiplets $Q^i_a$;  $a,b$ are
$\SU(2)$ color indices.  One can show \cite{cdsw0211,s0212} that a 
complete basis of local gauge-invariant operators in the chiral ring 
in this theory is comprised of just the glueball $S\sim W_\a\cdot W^\a$ 
and the meson operators $M^{ij}\sim Q^i\cdot Q^j$.  At a suitably
symmetric vacuum, say $\langle Q^i_a \rangle = \m \d^i_a$, the gauge 
bosons and the quarks $Q^i_a$ with $i>2$, as well as their
superpartners, all get mass $\m$ by the Higgs mechanism.  So, since 
the glueball and meson operators only involve the product of two 
fundamental fields, they create modes of particle states with mass 
at most $2\m$.  (The masses just add since, by taking $\m<<\L$, we 
are at arbitrarily weak coupling.)  Thus in a Wilsonian effective 
action found by integrating out modes above $2\m$ we may consistently 
keep all components of $S$ and $M^{ij}$, and since they generate the
chiral ring, there must exist an effective superpotential which is a 
function of only these chiral fields.

So far we have argued that an effective superpotential for local 
gauge-invariant operators in the chiral ring exists and makes 
sense for superQCD with enough massless flavors that it is IR free. 
This does not show the existence of such an effective superpotential 
in the asymptotically free case.  In particular, for theories in the
``conformal window" where neither the direct nor Seiberg dual 
description is IR free \cite{s9411} (\eg\ $3<\nf<6$ for $\SU(2)$ 
gauge group), we have no useful description of the light degrees of
freedom at the origin of moduli space.  Nevertheless, given an 
effective superpotential for an IR free theory, we can always 
integrate out flavors using holomorphy to derive consistent 
effective superpotentials in the conformal window.  
This round-about argument assures us that effective superpotentials 
exist for all numbers of light flavors in superQCD.
 
\subsection{Outline of the paper}

In this paper we illustrate this line of reasoning for the simplest 
example: four-dimensional $N=1$ supersymmetric $\SU(2)$ QCD with many 
light fundamental flavors.  The form of the effective superpotential 
is fixed by the global symmetries, making this a particularly easy 
case to study.  

\FIGURE{
\epsfig{file=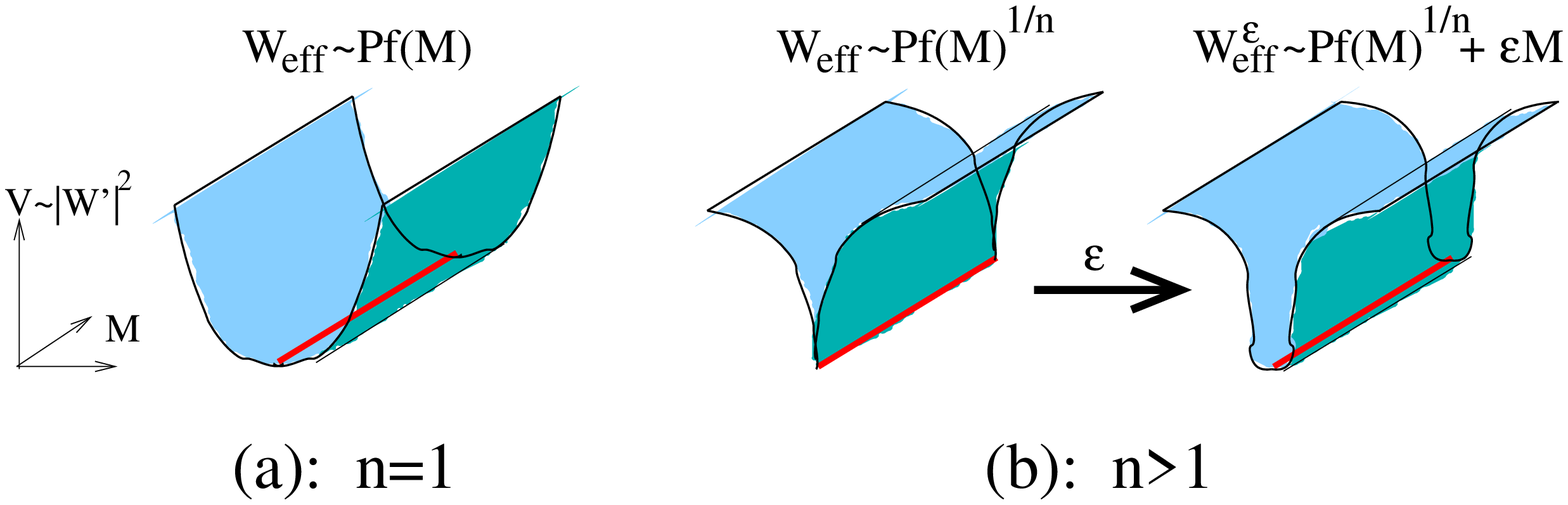,width=35em}		
\caption{Sketches of the effective potential as a function of
the meson vevs for $\SU(2)$ superQCD with (a) $n:=\nf-2=1$,
where the potential is regular, and (b) for $n>1$, 
where the potential has a cusp but can be smoothed by 
a small perturbation $\ve$.  Red lines denote the moduli spaces 
(vacua).}}

We start in section 2 by assuming we can integrate out the glueball
degrees of freedom to express the superpotential solely in terms
of the meson vevs.  The resulting superpotentials, determined by
the symmetries, are singular.  We show that they are, nevertheless, 
perfectly sensible.  The cusp-like behavior of their associated 
potentials still unambiguously describes their supersymmetric 
minima.  They can be regularized by turning on arbitrarily small 
quark masses.  We then observe that no matter how the masses are 
sent to zero, these superpotentials always give the correct 
constraint equation describing the moduli space.  The basic point 
is illustrated in figure 1: even though the potential has cusp-like
singularities all along the moduli space, it nevertheless has
a well-defined minimum.  We also show that upon giving large
masses to some flavors and integrating them out, we recover 
the superpotential for fewer numbers of flavors. 

In section 3 we justify the assumption that the glueball field
can be consistently integrated out.  Using the Konishi anomaly
\cite{k84,ks85}, one can derive a partial differential 
equation satisfied by the superpotential as a function of the 
meson and glueball vevs.  We solve these equations, determining
the integration function by matching to the Veneziano-Yankielowicz
superpotential \cite{vy82} for pure $\SU(2)$ superYang-Mills.  
Since in the IR free case we have included all the local chiral 
light degrees of freedom, by the arguments of this section
we expect these differential equations to be integrable and the 
superpotential to exist.  Indeed they are and it does, and matches
(upon integrating out the glueballs) the results of section 2.

In the asymptotically free cases in the conformal window, $3<\nf<6$,
since there is no argument that it is consistent to describe the
effective theory in terms of the local gauge-invariant chiral ring 
made from the microscopic fields, it is possible that the differential
equations for the effective superpotential derived from the Konishi
anomaly may not be integrable.  In the case of $\SU(2)$ superQCD, 
however, we find that they are integrable.  This is presumably an
``accident" due to the large global symmetry group of the theory, and
need not remain the case for $\SU(\nc)$ with $\nc>2$ \cite{ae}.
We also check that we get the same superpotential by using the Konishi 
anomaly equation in both the direct and Seiberg dual 
description of the low energy theory. 

We justified the existence of these singular effective 
superpotentials in IR free theories.  
By integrating out flavors we can use them to 
deduce the correct effective superpotentials for few numbers of 
flavors where quantum effects dramatically alter the form of the 
superpotential (first deforming the classical moduli space, then 
lifting it altogether as flavors are integrated out).  It was 
shown in \cite{bw0409} that in a 
description in terms of only the massless multiplets in the vicinity 
of an arbitrary non-singular point on the moduli space, these strong 
quantum effects descend from higher-derivative F-terms which can be 
calculated using instanton methods.  It is therefore a non-trivial, 
and quite elaborate, check of our singular superpotentials that by
expanding them around a generic vacuum and integrating out at tree 
level the massive modes of the meson field (those that take us off 
the moduli space), we reproduce the higher-derivative F-terms
computed in \cite{bw0409}.  We perform this check in section 4.

Singular superpotentials are a generic feature of gauge theories with
a large number of flavors, and are not special just to four-dimensional
theories.  In section 5 we argue in an example with three-dimensional 
$N=2$ supersymmetry where the global symmetries are enough to fix a 
singular form for the effective superpotential, that these superpotentials
satisfy a similar set of consistency checks as do the four-dimensional 
theories.  However, in this case we no longer have an IR free regime 
as a starting point from which to derive effective superpotentials 
by integrating out flavors using holomorphy.  Thus the meaning of 
singular effective superpotentials is less certain in $d<4$.

\section{Effective superpotentials for SU(2) superQCD with $\bf\nf>3$} 

$N=1$ supersymmetric $\SU(2)$ QCD has an adjoint vector multiplet 
$W^{(ab)}_\a$ containing the gluons and $2\nf$ massless quark chiral 
multiplets $Q^i_a$ in the fundamental representation; $i,j=1,\ldots,
2\nf$ are flavor indices and $a,b=1,2$ are $\SU(2)$ color indices.  
(There must be an even number of flavors for anomaly cancellation 
\cite{w82}.)  The classical 
moduli space of vacua is conveniently parametrized in terms of the vevs 
of the antisymmetric, gauge-singlet chiral meson fields $\h M^{[ij]} := 
Q^i_a \e^{ab} Q^j_b $, where $\e^{ab}$ is the invariant antisymmetric 
tensor of $\SU(2)$.  For $\nf=1$ the classical moduli space is the 
space of arbitrary meson vevs $M^{ij}$, while for $\nf\ge2$ it is all 
$M^{ij}$ satisfying the constraint
\be\label{2.1}
\e_{i_1\cdots i_{2\nf}} M^{i_1 i_2} M^{i_3 i_4} = 0,
\ee
or, equivalently, rank$(M)\leq2$.

The moduli space is modified by quantum effects when $\nf<3$.  For 
$\nf=1$, there is a dynamically generated superpotential which lifts 
all the classical flat directions \cite{ads84},
\be\label{2.2}
\we ={\L^5\over\Pf M} ,\qquad\qquad\qquad (\nf=1)
\ee
where $\L$ is the strong-coupling scale of the theory and
the Pfaffian is defined as $\Pf M := \e_{i_1\cdots i_{2\nf}}
M^{i_1i_2}\cdots M^{i_{2\nf-1}i_{2\nf}}=\sqrt{\det M}$.
For $\nf=2$ the superpotential can be written \cite{s9402}
\be\label{2.3}
\we=\Sigma \left(\Pf M - \L^4\right) ,\qquad (\nf=2)
\ee
where $\Sigma$ is a Lagrange multiplier enforcing a quantum-deformed
constraint $\Pf M = \L^4$, which removes the singularity at the origin 
of the classical moduli space.

For $\nf=3$ the superpotential is \cite{s9402}
\be\label{2.4}
\we=-{1\over \L^3}\Pf M.  \qquad\qquad (\nf=3)
\ee
The resulting equations of motion reproduce the classical constraint 
(\ref{2.1}), which are therefore not modified quantum mechanically.  
Note that although the superpotential (\ref{2.4}) apparently diverges 
in the weak-coupling $\L\to0$ limit, it actually vanishes on the moduli 
space since (\ref{2.1}) implies $\Pf M=0$.  The negative power of $\L$ 
reflects the fact that fluctuations off the classical constraint surface 
become infinitely massive in the weak coupling limit. 

\subsection{Singular superpotential for $\bf\nf>3$ and the classical 
constraints}

For $\nf>3$, the classical constraints are also not modified quantum
mechanically.  However, the complex singularities of the moduli space 
defined by (\ref{2.1}) indicate the presence of new massless degrees 
of freedom there, in addition to the components of $\h M^{ij}$ 
\cite{s9411}.

We argued in section 1 that, nevertheless, an effective 
superpotential for the IR free case ($\nf>5$) should exist as a 
function\footnote{It need not be single valued; it is allowed to 
shift by integral multiples of $2\pi i S$, reflecting the angularity 
of the theta angle.} of the unconstrained chiral meson and glueball 
vevs, $M^{ij}$ and $S$. 
For the moment, let us assume that the glueball superfield
can always be conistently
integrated out away from the origin, so we can just deal with an 
effective superpotential depending only on $M^{ij}$.  Then the possible
form of the effective superpotential is completely determined by the
symmetries up to an overall numerical factor.

The only effective superpotential consistent with holomorphicity, 
weak-coupling limits, and the global symmetries is \cite{ip9505}
\be\label{spweff}
\we=-n\left(\Pf M\over \L^{b_0} \right)^{1/n},
\qquad
n:= \nf-2 >1
\ee
where $b_0 = 6-\nf$ is the coefficient of the one-loop beta function.  
The coefficient in (\ref{spweff}) will be justified below.  We will
also check below that this superpotential is consistent under integrating
out successive flavors, and so its form in the asymptotically free
cases ($\nf<6$) follow from any IR free case ($\nf\ge6$) by holomorphy
and RG flow.  We leave the justification of the assumption that $S$
can be integrated out to section 3.

The fractional power of $\Pf M$ in (\ref{spweff})
implies that this superpotential has a cusp-like singularity at its
extrema.  The rest of this paper is devoted to 
arguing that this superpotential is nevertheless correct.

The first issue is how the classical constraint (\ref{2.1}) follows
from extremizing (\ref{spweff}).  Because these superpotentials are 
singular at their extrema, we cannot just take derivatives.  Instead, 
we deform $\we$ by introducing regularizing parameters before extremizing.
Independent of how the regularizing parameters are sent to 
zero, the extrema of the superpotential will give the classical
constraints (\ref{2.1}). 

We regularize (\ref{spweff}) by adding a mass term with an invertible 
antisymmetric mass matrix $\ve_{ij}$ for the meson fields,
\be\label{regweff}
\wep := \we + {1\over 2} \ve_{ij}M^{ij}.
\ee
We regularize $\we$ with a term linear in $M$ since that is the only
integral power of $M$ that smoothes the minima but is subleading at
large $M$, and therefore does not create spurious extrema.
Varying $\wep$ with respect to $M^{kl}$ yields the equation of motion
\be
M^{kl}=-\L^{-b_0/n}(\Pf M)^{1/n}(\ve^{-1})^{kl}.
\ee  
Solving for 
$\Pf M$ in terms of $\ve$ and substituting back gives 
$M^{kl}=-\L^{b_0/2}(\Pf \ve)^{1/2}(\ve^{-1})^{kl}$
which in turn implies
\be\label{2.19}
\e_{i_1\ldots i_{2\nf}} M^{i_1 i_2} M^{i_3 i_4} 
= {1\over\L^{b_0}} \e_{i_1\ldots i_{2\nf}}
(\ve^{-1})^{i_1 i_2} (\ve^{-1})^{i_3 i_4}\ \Pf \ve.
\ee
The right hand side of the above expression is a polynomial 
of order $n > 0$ in the $\ve_{ij}$.  Therefore, no matter how we send
$\ve_{ij}\to 0$, the right hand side will vanish, giving back the
classical constraint (\ref{2.1}).  Furthermore, it is easy to check that
any solution of the classical constraint can be reached in this way.

It may be helpful to present another, less formal, way of seeing how 
the classical constraint emerges from the singular effective 
superpotential.
Use the global symmetry to rotate the meson fields into the skew 
diagonal form  
\bea\label{2.21}
M^{ij}&=&\pmatrix{M_1& & & \cr\ &M_2& & \cr\ & & \ddots & \cr\ &
& &M_{\nf}\cr}\otimes i\s_2,
\eea
so the effective superpotential (\ref{spweff}) becomes 
\be
\we = -n \L^{-b_0/n} \left(\prod_i M_i\right)^{1/n}.
\ee  
The equations of motion which follow from extremizing 
with respect to the $M_i$ are
\be\label{2.23}
M_i^{{1\over n}-1}  \prod_{j\neq i} M_j^{1\over n}=0.
\ee
Though these equations are ill-defined if we set any of the $M_i=0$, 
we can probe the solutions by taking limits as some of the $M_i$ 
approach zero.  To test whether there is a limiting solution where 
$K$ of the $M_i$ vanish, consider the limit $\ve\to0$ with 
$M_1\sim\ve^{\a_1}, \ldots, M_K\sim\ve^{\a_K}$ with $\a_j>0$ to be 
determined. Plugging into 
(\ref{2.23}), only the first $K$ equations have non-trivial limits
\be\label{2.24}
\lim_{\ve\to0}\ve^{{1\over n} \left(\sum_j \a_j\right)-\a_i}=0,
\qquad i=1,\ldots, K,
\ee
giving the system of inequalities $n\a_i< \sum_j \a_j$ for $i=1,\ldots,K$.  
These inequalities have solutions if and only if $K>n$, implying that 
rank$(M)\leq 2$ which is precisely the classical constraint (\ref{2.1}).

Note that, as in the $\nf=3$ case discussed above, the
negative power of $\L$ appearing in the effective superpotential
(\ref{spweff}) is not inconsistent with the weak coupling limit
because the equations of motion (\ref{2.1}) following from the
superpotential imply $\Pf M=0$, so that (\ref{spweff}) vanishes
on the moduli space.  

\subsection{Consistency upon integrating out flavors}

Besides correctly describing the moduli space, the effective
superpotentials should also pass some other tests.  If we add a 
mass term for one flavor in the superpotential of a theory with 
$\nf$ flavors and then integrate it out, we should recover the 
superpotential of the theory with $\nf-1$ flavors.  To show that 
the effective superpotential (\ref{spweff}) passes this test, we 
add a gauge-invariant mass term for one flavor, say $M^{{2\nf-1}~
{2\nf}}$,
\be\label{2.26}
\we =-n\L^{-b_0/n} (\Pf M)^{1/n}+ mM^{{2\nf-1}~{2\nf}}.
\ee  
The equations of motion for $M^{{i}~{2\nf-1}}$ and $M^{{j}~{2\nf}}$ 
(for $i\neq 2\nf-1$ and $j\neq 2\nf$) put the meson matrix into the 
form $M^{ij} = {\wh M \  0 \choose 0 \ \wh X}$ where $\wh M$ is a 
$2(\nf-1)\times 2(\nf-1)$ and $\wh X$ a $2\times2$ matrix.  
Integrating out $\wh X\sim M^{{2\nf-1}~{2\nf}}\otimes \s_2$ 
by its equation of motion gives
\be\label{2.28}
\we = -(n-1)\wh\L^{b_0/(n-1)} (\Pf \wh M)^{1/(n-1)},
\ee 
where $\wh\L = m \L^{6-\nf}$ is the strong-coupling scale 
of the theory with $\nf-1$ flavors, consistent with matching the 
RG flow of the couplings at the scale $m$.  Dropping the hats, we 
recognize (\ref{2.28}) as the effective superpotentials of $\SU(2)$ 
superQCD with $\nf-1$ flavors.

\section{Consistency with the Konishi anomaly equation}

The Konishi anomaly implies a differential equation which the
effective superpotential should obey when considered as a function 
of the meson and glueball vevs.  We outline here the derivation of 
this equation and show that its solution enables us to determine 
the dependence of the effective superpotential on the glueball vev, 
and to justify the assumption that made in section 2 that the 
glueball superfield can be consistently integrated out.  Although 
this is a simple exercise, it gains interest when compared to the 
$\SU(N)$ case where the corresponding generalized Konishi anomaly 
equations \cite{cdsw0211} are much more complicated \cite{c0305,ae}, as 
mentioned in section 1.  

In the chiral ring the Konishi anomaly \cite{k84,ks85}
for a tree level superpotential $W_{\rm tree}$ takes the form 
\be\label{kadir1}
\langle {\del W_{\rm tree}\over \del Q_a^i} Q_a^j\rangle = S \d^j_i.
\ee
where $S$ is the vev of the glueball superfield 
$\h S={1\over {32\pi^2}}\tr ({W^\a W_\a})$.
(We distinguish an operator from its vev by putting a hat on the operator.) 
This is a special case of the generalized Konishi anomaly, which is
perturbatively one-loop exact \cite{cdsw0211}, and has also been shown 
\cite{s0311} to be non-perturbatively exact for a $\U(N)$ gauge 
theory with matter in the adjoint representation as well as for $\Sp(N)$ 
and $\SO(N)$ gauge theories with matter in symmetric or antisymmetric 
representations.  For the theory we are discussing here, we will not prove 
that the Konishi anomaly is non-perturbatively exact, though presumably
this can be done along the lines of \cite{s0311}.  Instead, because the
global symmetry of the $\SU(2)$ superQCD uniquely determines the 
superpotential as discussed in the previous section, we only need
check that the Konishi anomaly equation implies this form of the 
superpotential.  This check serves as evidence for the non-perturbative
exactness of the Konishi anomaly equation for the theory under discussion.
Had the Konishi anomaly equation been modified non-perturbatively, we would 
have found a different result for $\we$ in this section.

\subsection{Direct description}


In the Konishi anomaly equation (\ref{kadir1}), take as our tree level superpotential  
\be\label{kadir2}
W_{\rm tree}= m_{ij}(\h M^{ij}-M^{ij}),
\ee
so that
\be\label{kadir3}
m_{ij}=-{1\over 2}{{\del \we} \over  \del M^{ij}},
\ee
is a Lagrange multiplier imposing that $M^{ij}$ are the vacuum 
expectation values of the meson operators $\h M^{ij}$. 
Substituting (\ref{kadir2}) into (\ref{kadir1}) and using the fact 
that the expectation value of a product of gauge-invariant chiral 
operators equals the product of the expectation values of the individual 
ones, gives $2 m_{ik} M^{kj} = S \d^j_i$.  Using (\ref{kadir3}) we then 
obtain a partial differential equation for the effective superpotential,
\be\label{kadir4}
{\del \we \over \del M^{ik}} M^{kj} = S \d^j_i,
\ee
whose solution is
\be\label{kadir5}
\we(M, S)= S \ln \left(\Pf M \over \L^{2\nf}\right)+ f(S), 
\ee
where $f(S)$ is an undetermined function.  Upon giving the quarks a 
mass $m$ and integrating them out, the superpotential reduces to 
$f(S)+\nf S [\ln S-\ln(m\L^2)-1]$.  In the limit $m\to\infty$, $\L\to0$ 
keeping $\L_0$ fixed, where $6\ln\L_0=b_0\ln\L+\nf\ln m$, 
this becomes the $\SU(2)$ superYang-Mills theory with strong 
coupling scale $\L_0$.  The superpotential for this theory is the 
Veneziano-Yankielowicz superpotential \cite{vy82} $W_{\rm VY}(S) = 
2S\left[\ln (S/\L_0^3) -1\right]$, implying that
\be\label{kadir6}
f(S) =  (2-\nf)S\left[\ln (S/\L^3) - 1\right].
\ee
Substituting (\ref{kadir6}) into (\ref{kadir5}) gives the effective
superpotential as a function of $S$ and $M^{ij}$.  It is easy to
see that at its extrema $S$ is massive (except at the origin), 
justifying the assumption of the last section that it could be 
integrated out.  Finally, integrating $S$ out by solving its equation of 
motion, we arrive at the effective superpotential (\ref{spweff}).

\subsection{Seiberg dual description}

Viewing our $\SU(2)$ theory as an $\Sp(1)$ gauge theory, when $\nf>3$ 
the theory has a Seiberg dual description \cite{ip9505} in terms of an 
$\Sp(\nf-3)$ gauge group.\footnote{The $\SU(\nf-2)$ Seiberg dual 
description \cite{s9411} is more difficult to analyze since it has 
a smaller global symmetry group.}  The dual $\Sp(\nf-3)$ theory has 
$2\nf$ dual quark chiral multiplets $q^a_i$ in the fundamental 
representation as well as a gauge-singlet chiral multiplet $\h\MM^{[ij]}$ 
which is coupled to the dual meson fields $\h\NN_{ij}:=q^a_i J_{ab} q^b_j$ 
through the superpotential $W=\h\NN_{ij}\h\MM^{ij}$.  Here $J_{ab}$ is
the invariant symplectic antisymmetric tensor, $i,j=1,\ldots,2\nf$ are
flavor indices, and $a,b=1,\ldots,2\nf-6$ are the gauge indices.  This superpotential gives masses to the dual quarks and sets $\NN_{ij}=0$ 
when $\MM^{ij}\neq0$.  The dual description is IR free when $\nf<6$.

To determine the effective superpotentials of the dual theory 
we can either use the global symmetry, weak-coupling limit and 
the holomorphicity argument, or the Konishi anomaly equations. 
Both give the same answer;  we discuss the Konishi anomaly
equations.  The ring of local gauge-invariant chiral operators 
is generated by $\h\SS$, $\h\MM^{ij}$ and $\h\NN_{ij}$ \cite{w0302}. 
The Konishi anomaly equations are $\langle q^a_j(\del W_{\rm tree}
/\del q^a_i) \rangle = \SS \d^i_j$.  Take as the tree level superpotential
\be\label{kasd2}
W_{\rm tree} = \h\NN_{ij}\h\MM^{ij}+m_{ij}(\h\MM^{ij}-\MM^{ij}),
\ee
so that as before, $m_{ij}=-{1\over2}(\del\we/\del\MM^{ij})$,
is a Lagrange multiplier imposing that $\MM^{ij}$ are the
vacuum expectation values of the scalar operators $\h\MM^{ij}$. 
We have not included a Lagrange multiplier for the dual mesons
$\h\NN_{ij}$ because our analysis is valid only for points away 
from the origin of the moduli space where the dual quarks are massive. 

As in the direct description, the Konishi anomaly with (\ref{kasd2}) gives
$2\MM^{ik}\NN_{kj}=-\SS\d^i_j$.  The $\h\MM^{ij}$ equation of motion 
gives $\NN_{ij}=-m_{ij}$, giving the partial differential equation 
$\MM^{ik}(\del\we/\del\MM^{kj})=\SS\d^i_j$ whose solution is
\be\label{kasd3}
\we(\MM, \SS) = \SS \ln \left[{\Pf\MM \over \t\L^{\nf}}\right]+ f(\SS). 
\ee
$f(\SS)$ is determined as before to be $f(\SS) = (2-\nf) \SS 
[\ln(\SS/\t\L^3)-1]$.  Integrating out $\SS$ then gives the
effective superpotential in the dual description
\be\label{kasd4}
\we=(\nf-2){\left(\t\L^{2\nf-6}
\Pf\MM \right)}^{1\over{\nf-2}}.
\ee 
The dual and direct descriptions are equivalent in the IR; the 
$\MM^{ij}$ are identified with the direct theory mesons by $\MM^{ij} 
= {1\over\m} M^{ij}$, where $\m$ is a mass scale related to the 
dual and the direct theory strong-coupling scales by \cite{ip9505}
\be\label{kasd5}
\L^{6-\nf}\t\L^{2\nf-6}=(-1)^{\nf} \m^{\nf}.
\ee
Rewriting (\ref{kasd4}) in terms of $\L$ and $M^{ij}$ gives our
superpotential (\ref{spweff}).

\section{Higher-derivative F-terms}

In this section we show that the effective superpotential 
(\ref{spweff}) passes a different, more stringent, test.  In 
\cite{bw0409} a series of higher-derivative F-terms were 
calculated by integrating out massive modes at tree level from 
the non-singular effective superpotentials (\ref{2.3}) and (\ref{2.4}) 
for $\SU(2)$ superQCD with $\nf=2$ and $3$, and by an 
instanton calculation for $\nf\ge3$.   In this section
we show that our singular superpotential for $\nf>3$ 
reproduces these F-terms by a tree-level calculation.  As in 
our discussion of the classical constraint in the last section, the 
key point in this calculation is to first regularize the effective 
superpotential (\ref{spweff}), and then show that the results are 
independent of the regularization.  

The higher-derivative F-terms found in \cite{bw0409} in $\SU(2)$
superQCD are, for $\nf\ge2$ flavors,
\bea\label{bw}
\d S &=& \int d^4xd^2\th\,\L^{6-\nf}(M\Mb)^{-\nf}
\e^{i_1 j_1 \cdots i_\nf j_\nf}\Mb_{i_1 j_1} \nonumber\\
&&\quad\mbox{}\times
(M^{k_2 \ell_2} \Db\Mb_{i_2 k_2} \cdot \Db\Mb_{j_2 \ell_2}) \cdots 
(M^{k_\nf \ell_\nf} \Db\Mb_{i_\nf k_\nf} \cdot \Db\Mb_{j_\nf \ell_\nf}),
\eea
where $(M\Mb) := (1/2)\sum_{ij} M^{ij}\Mb_{ij}$, and the dot denotes 
contraction of the spinor indices on the covariant derivatives $\Db_\ad$.  
Although these terms are written in terms of the unconstrained meson 
field, they are to be understood as being evaluated on the classical 
moduli space.  In other words, we should expand the $M^{ij}$ in 
(\ref{bw}) about a given point on the moduli space, satisfying 
(\ref{2.1}), and keep only the massless modes (\ie\ those tangent 
to the moduli space).  This should be 
contrasted with our effective superpotential (\ref{spweff}) which 
makes sense only in terms of the unconstrained meson fields.

Note that even though (\ref{bw}) is written as an F-term (an integral 
over a chiral half of superspace), the integrand is not obviously a 
chiral superfield.  But the form of the integrand is special: it is 
in fact chiral, and cannot be written as $\Db^2$(something), at least 
globally on the moduli space, and so is a protected term in the low 
energy effective action.  These features of (\ref{bw}), discussed in 
detail in \cite{bw0409}, will neither play an important 
role nor be obvious in our derivation of these terms.

We will now show how (\ref{bw}) emerges from the effective superpotential 
(\ref{spweff}).  To derive effective interactions for massless modes 
locally on the moduli space from the effective superpotential for the 
unconstrained mesons, and which therefore lives off the moduli space,
we simply have to expand the effective superpotential around a given 
point on the moduli space and integrate out the massive modes at tree 
level.  The only technical complication is that, as discussed in section 
2, the effective superpotential needs to be regularized first, \eg\ by 
turning on a small mass parameter $\ve_{ij}$ as in (\ref{regweff}), so 
that it is smooth at its extrema.  At the end, we take $\ve_{ij}\to 0$.  
The absence of divergences as $\ve\to0$ is another check of the 
consistency of our singular effective superpotential.

\subsection{Taylor expansion around a vacuum}

The moduli space is defined by the constraint rank$(M)\le2$ (\ref{2.1}).  Without loss of generality, we can choose the vacuum satisfying 
(\ref{2.1}) around which we expand to be
\be\label{sp2vac}
\Mo^{ij}=\pmatrix{\m&&&\cr &0&&\cr &&\ddots&\cr &&&0\cr}
\otimes i\s_2,
\ee
with $\m$ a non-vanishing constant, by making an appropriate $\SU(2\nf)$ 
global flavor rotation.  Note that $\Mo^{ij}$ breaks the $\SU(2\nf)$ 
global symmetry to $\SU(2)\times\SU(2\nf-2)$.  Accordingly we henceforth 
partition the $i,j$ flavor indices into those transforming under the 
unbroken $\SU(2)$ factor from the front of the alphabet---$a, b {=} 1, 
2$---and the remaining $\SU(2\nf-2)$ indices from the back: $u, v,\ldots 
= 3, \ldots, 2\nf$.  Linearizing (\ref{2.1}) about (\ref{sp2vac}), 
$M^{ij} = \Mo^{ij} + \dM^{ij}$, implies that the massless modes are 
$\dM^{12}$ and $\dM^{au}$, while the $\dM^{uv}$ are all massive.  The 
$\dM^{12}$ mode can be absorbed in a rescaling of $\m$, so we only need 
to focus on the $\dM^{au}$ modes.

Expanding (\ref{bw}) around $\Mo^{ij}$ and keeping only the massless 
modes, we generate an infinite number of terms.  The leading term, 
which is of order $(\dMb)^{2\nf-2}$, reads
\bea\label{3.4}
\d S &\sim& \int d^4xd^2\th\, \L^{6-\nf} \mb^{1-\nf} \m^{-1} 
\e^{u_1v_1\cdots u_{\nf-1}v_{\nf-1}}
\,(\Db\dMb_{1u_1} \cdot \Db\dMb_{2v_1})\times \nonumber\\ 
&&\cdots\times (\Db\dMb_{1u_{\nf-1}} \cdot \Db\dMb_{2v_{\nf-1}}),
\eea
since $\Db\Mbo=0$.  It suffices to show that this leading term
is generated in perturbation theory since the $\SU(2\nf)$ flavor 
symmetry together with the chirality of the integrand imply that 
(\ref{bw}) is the unique non-linear completion of (\ref{3.4}); see 
section 3.2 of \cite{bw0409}.\footnote{We could, in principle, 
directly generate the higher-order terms in the expansion of 
(\ref{bw}) by a tree level calculation.  In fact, a sixth-order 
term in the $\nf=3$ theory is calculated in this way in \cite{bw0409}.}

In order to demonstrate how (\ref{3.4}) is generated at tree level
from our effective superpotential, we first regularize $\we\to\wep$, 
which we repeat here:
\be\label{regweff2}
\wep := -n\l (\Pf M)^{1/n} + {1\over 2}\ve_{ij} M^{ij},
\ee
where we have defined the convenient shorthands
\be
n:= \nf-2,\qquad\qquad
\l := \L^{-b_0/n}.
\ee
Now the extrema of $\wep$ no longer satisfy the classical constraint 
equation (\ref{2.1}) but are deformed as in (\ref{2.19}).  
So we must also deform (\ref{sp2vac}) as well.  It is convenient to 
choose $\ve_{ij} = \l \ve^{1/n} \m^{(1-n)/n} \mbox{diag} 
\{\ve,\m,\ldots,\m\} \otimes i\s_2$ so that
\be\label{3.3}
(\Mvo)^{ij}=\pmatrix{\m&&&\cr &\ve&&\cr &&\ddots&\cr&&&\ve\cr}
\otimes i\s_2.
\ee
An advantage of this choice is that it preserves an $\SU(2) \times 
\Sp(2\nf-2)$ subgroup of the flavor symmetry.  In the limit $\ve\to0$ 
this is enhanced to $\SU(2)\times\SU(2\nf-2)$.  Also, the massless 
directions around this choice are still $\dM^{ua}$ as before.

\subsection{Feynman rules}

We use standard superspace Feynman rules \cite{ggrs0108} to compute 
the leading interaction term in the effective action for the massless 
$\dM^{ua}$ modes by integrating 
out the massive $\dM^{uv}$ modes.  This means we need to evaluate 
connected tree diagrams at zero momentum with internal massive 
propagators and external massless legs.  The massive modes have standard 
chiral, anti-chiral, and mixed superspace propagators with masses 
derived from the quadratic terms in the expansion of $\wep$.  The 
higher-order terms in the expansion give chiral and anti-chiral 
vertices.  

A quadratic term in the superpotential, $W={1\over2}m(\dM)^2+\cdots$,
gives a mass which enters the chiral propagator as
$\langle\dM\dM\rangle = \bar m (p^2+|m|^2)^{-1} (D^2/p^2)$,
similarly for the anti-chiral propagator, and as
$\langle\dM\dMb\rangle = (p^2+|m|^2)^{-1}$ for the mixed
propagator.  Each propagator comes with a factor of $\d^4(\th -\th')$.
Even though the diagrams will be evaluated at zero momentum, we must 
keep the $p^2$-dependence in the above propagators for two reasons.  
First, there are spurious poles at $p^2=0$ in the (anti-)chiral 
propagators which will always cancel against momentum dependence 
in the numerator coming from $\Db^2$'s in the propagators and 
$D^2$'s in the vertices.  For instance, $D^2\Db^2=p^2$ when acting 
on an anti-chiral field, giving a factor of $p^2$ in the numerator 
which can cancel that in the denominator of the anti-chiral propagator, 
to give an IR-finite answer.  Second, expanding the IR-finite parts 
in a power series in $p^2$ around $p^2=0$ can give potential 
higher-derivative terms in the effective action, when $p^2$'s act 
on the external background fields.

Expanding $\wep$ around $(\Mvo)^{ij}$ gives the quadratic terms
\be\label{3.5}
\wep (\Mvo+\dM) = \wep(\Mvo) + \l t^{ijk\ell}_{i'j'k'\ell'}
(\Pf\Mvo)^{1/n} (\Mvo)^{-1}_{ij}(\Mvo)^{-1}_{k\ell} 
\dM^{i'j'}\dM^{k'\ell'}
+\cdots .
\ee
We will drop for now the numerical tensor $t^{ijk\ell}_{i'j'k'\ell'}$ 
which controls how the $ij\ldots$ indices are contracted with the
$i'j'\ldots$ indices, though its form will be needed for a later
argument.  But for our immediate purposes, it suffices to note,
as we discuss below, that in the $\ve\to0$ limit the tensor structure 
of our tree diagrams is fixed by the $\SU(2)\times \SU(2\nf-2)$ 
subgroup of the global symmetry that is preserved by the vacuum.

Specializing to the massive modes for which 
$\{i,j,k,\ell\}\to\{u,v,w,x\}$, and using (\ref{3.3}), 
then gives the mass $m\sim \l\ve^{-\a}\m^\b$ where
\be\label{abeq}
\a:={n-1\over n}, \qquad\qquad
\b:={1\over n} .
\ee
The propagators are then
\bea\label{props}
\dM^{uv}\ \hbox{\bf--\, --\, --\, --}\ \dM^{wx} \quad &\sim &\quad 
{\ve^\a\over\l\m^\b} {D^2 \over p^2} 
\left(1+\left|\ve^\a\over\l\m^\b\right|^2 p^2\right)^{-1},
\nonumber\\
\dMb_{uv}\ \hbox{\bf-----------}\ \dMb_{wx} \quad &\sim &\quad 
{\veb^\a\over\lb\mb^\b} {\Db^2 \over p^2} 
\left(1+\left|\ve^\a\over\l\m^\b\right|^2 p^2\right)^{-1},
\nonumber\\
\dMb_{uv}\ \hbox{\bf-----\, --\, --}\ \dM^{wx} \quad &\sim &\quad 
\left|\ve^\a\over\l\m^\b\right|^2
\left(1+\left|\ve^\a\over\l\m^\b\right|^2 p^2\right)^{-1}.
\eea
We have suppressed the tensor structure on the $\{u,v,w,x\}$ indices.

The (anti-)chiral vertices come from higher-order terms in
the expansion of $\wep$ ($\bar\wep$).  Each (anti-)chiral  
vertex will have a $\Db^2$ ($D^2$) acting on all but one of its 
internal legs.  Also, each vertex is accompanied by an $\int d^4\th$.
The $\ell$th-order term in the expansion of $\wep$ has the general 
structure 
\be\label{3.6}
\l (\Pf\Mvo)^{1/n} (\Mvo)^{-1}_{i_1j_1}\cdots 
(\Mvo)^{-1}_{i_\ell j_\ell}\dM^{i_1'j_1'}\cdots \dM^{i_\ell'j_\ell'},
\ee
where we have suppressed the tensor structure which governs 
the order in which the $i'j'$ indices are contracted with the $ij$ 
indices.  Thus vertices with $m$ massless legs and $\ell-m$ massive 
legs are accompanied by the factors
\be\label{3.8}
\underbrace{\overbrace{
\lower20pt\hbox{\epsfxsize=4em\epsfbox{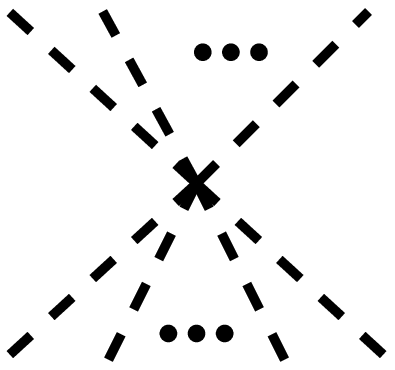}} 	
}^{m\ \mbox{\footnotesize{massless}}}}_{\ell-m\ 
\mbox{\footnotesize{massive}}}
\quad\sim\quad {\l\over\ve^{\g_{\ell,m}}\m^{\k_m}}, \qquad \qquad
\underbrace{\overbrace{
\lower20pt\hbox{\epsfxsize=4em\epsfbox{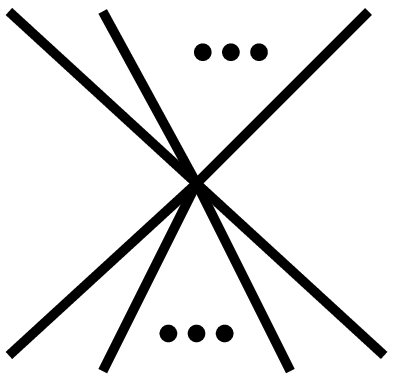}} 	
}^{m\ \mbox{\footnotesize{massless}}}}_{\ell-m\ 
\mbox{\footnotesize{massive}}}
\quad\sim\quad {\lb\over\veb^{\g_{\ell,m}}\mb^{\k_m}},
\ee
where
\be\label{gkeq}
\g_{\ell,m} := \ell-{m\over2}-{n+1\over n},
\qquad\qquad
\k_m := {m\over2}-{1\over n}.
\ee
Note that it follows from (\ref{3.6}) that the number $m$ of 
massless legs $\dM^{au}$ must be even, and furthermore half 
must be $\dM^{1u}$'s and half $\dM^{2u}$'s.  This is because 
these legs each have one index $a\in\{1,2\}$ and the only 
non-vanishing components of $(\Mvo)^{-1}_{ij}$ with indices 
in this range are $(\Mvo)^{-1}_{12} = -(\Mvo)^{-1}_{21} = 
\m^{-1}$ which have two of these indices.

Finally, to each (anti-)chiral external leg at zero momentum
is assigned a factor of the (anti-)chiral background field
$\dM^{au}(x,\th)$ ($\dMb^{au}(x,\thb)$) all at the same $x$.  
Overall momentum conservation means that the diagram has a 
factor of $\int d^4x$.  The $\d^4(\th-\th')$ for each internal 
propagator together with the $\int d^4\th$ integrals at each 
vertex leave just one overall $\int d^4\th$ for the diagrams.

\subsection{$\bf\nf=3$}

We start by first looking at the $\nf=3$ case.  Although
this case does not involve a singular superpotential, it has 
the virtue of being simple and yet still illustrates how the 
potential IR poles cancel, and may help make the use of the 
Feynman rules clearer to the bewildered reader.  Also, although 
in \cite{bw0409} a tree diagram is computed for $\nf=3$, it 
is a $\dMb^6$ term (which was useful for comparing to an 
instanton computation) and not the leading $\dMb^4$ term 
which we will be computing.

The $\nf=3$ case is special since it can only involve anti-chiral
vertices.  There are two diagrams that contribute, shown in figure 
2a.  The first diagram, consisting of just an amputated 4-vertex 
with massless legs, vanishes.  This can be seen by a symmetry
argument; since the diagram comes with no powers of $\ve$, in the 
$\ve\to0$ limit its index structure must be, by the unbroken $\SU(4)$ 
part of the flavor symmetry, proportional to $\e^{u_1v_1u_2v_2} 
\dMb_{1u_1} \dMb_{2v_1} \dMb_{1u_2} \dMb_{2v_2}$.  Because there 
are no derivatives acting on the $\dMb$'s, this vanishes under the 
antisymmetrization of the $u_i$ or $v_i$ since the $\dMb$'s are 
bosons.  Alternatively, it is easy to calculate the index structure
of the 4-vertex directly by expanding $\wep$ directly as in (\ref{3.5}).

\FIGURE{
\epsfig{file=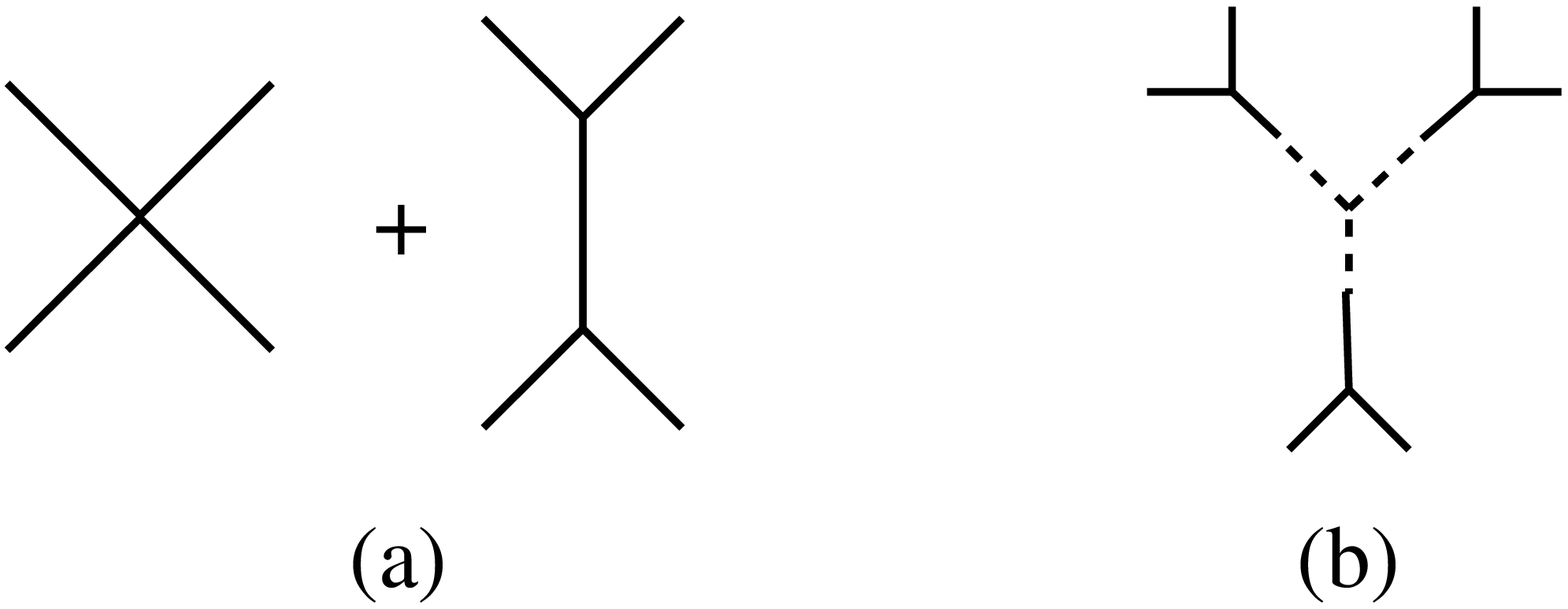,width=25em}     
\caption{Diagrams for (a) $\nf=3$, and (b) $\nf=4$.}}

Thus only the second diagram in figure 2a contributes.  Actually, two 
diagrams like this contribute: the one shown, and one in a crossed 
channel.  (The third channel does not contribute because, as noted
above, the 3-vertex with two external legs of the form
$\dMb_{1u} \dMb_{1v}$ or $\dMb_{2u}\dMb_{2v}$ vanishes by
antisymmetry.)  We will evaluate just one channel; the second gives
an identical result.  The Feynman rules give for the amplitude

\bea
{\cal A}_{\nf=3} &\sim& \int d^4xd^4\th_1d^4\th_2\ 
\dMb_{1u}(\th_1)\dMb_{2v}(\th_1) (J^{us}J^{vt}-J^{ut}J^{vs})
{\lb\over\veb^{\g_{3,2}}\mb^{\k_2}}\nonumber\\
&&\ \ \mbox{}\times \d^4(\th_1-\th_2) (J_{sp}J_{tq}-J_{sq}J_{tp})
{\veb^\a\over\lb\mb^\b}{\Db^2\over p^2} 
\left[1+\left|\ve^\a\over\l\m^\b\right|^2p^2\right]^{-1}\nonumber\\
&&\ \ \mbox{}\times {\lb\over\veb^{\g_{3,2}}\mb^{\k_2}}
(J^{wp}J^{xq}-J^{wq}J^{xp}) \dMb_{1w}(\th_2)\dMb_{2x}(\th_2)
\nonumber\\
&=& {\lb\over\mb}\e^{uvwx}\int d^4xd^4\th\ 
\dMb_{1u}\dMb_{2v} {\Db^2\over p^2}\left[1-|\m\l|^{-2}p^2
+{\cal O}(p^4)\right] \dMb_{1w}\dMb_{2x}
\nonumber\\
&=& {\lb\over\mb}\e^{uvwx}\int d^4xd^4\th\ 
\dMb_{1u}\dMb_{2v} \left[{\Db^2\over p^2}(\dMb_{1w}\dMb_{2x})
-|\m\l|^{-2}\Db^2(\dMb_{1w}\dMb_{2x})+{\cal O}(p^2)\right]
\nonumber\\
&=& {\lb\over\mb}\e^{uvwx}\int d^4xd^2\thb\ 
\dMb_{1u}\dMb_{2v} {D^2\Db^2\over p^2}(\dMb_{1w}\dMb_{2x})\nonumber\\
&&\ \ \mbox{}-{\e^{uvwx}\over\l\m\mb^2}\int d^4xd^2\th\ \Db^2[
\dMb_{1u}\dMb_{2v}\Db^2(\dMb_{1w}\dMb_{2x})]+{\cal O}(p^2)
\nonumber\\
&=& {\lb\over\mb}\e^{uvwx}\int d^4xd^2\thb\ 
\dMb_{1u}\dMb_{2v}\dMb_{1w}\dMb_{2x} 
\nonumber\\
&&\ \ \mbox{}-{\e^{uvwx}\over\l\m\mb^2}\int d^4xd^2\th\ 
(\Db\dMb_{1u}\cdot\Db\dMb_{2v})(\Db\dMb_{1w}\cdot\Db\dMb_{2x})
+{\cal O}(p^2) .
\eea
The first line includes the tensor structure of the vertices and 
propagator calculated by Taylor expanding $\wep$ around $(\Mvo)^{ij}$
as in (\ref{3.5}).  The antisymmetric symplectic tensor $J_{uv}$ 
and its inverse $J^{uv}$ arise from the structure of $\Mvo$ in 
(\ref{3.3});  it is simply $J:=\id_{\nf-1}\otimes i\s_2$, where
$\id_{\nf-1}$ is the $(\nf-1)\times(\nf-1)$ identity.  The 
second line performs a $d^4\th$ integration, the tensor algebra,
the Taylor expansion of the propagator around $p^2=0$, and substitutes
the $\nf=3$ values $\a=0$, $\b=-1$, $\g_{3,2}=0$, and $\k_2=0$ from
(\ref{abeq}) and (\ref{gkeq}).  The fourth line trades an $\int d^2\th$
for a $D^2$ in the first term, and a $\int d^2\thb$ for a $\Db^2$ in
the second term.  The fifth line uses the identity $D^2\Db^2=p^2$
on anti-chiral fields to cancel the IR pole in the first term, and
uses the equation of motion $\Db\dMb=0$ to leading order in $\dMb$
to distribute the $\Db$'s in the second term.  The first term in the 
last line cancels by antisymmetry, leaving the second term which is the 
higher-derivative F-term predicted in \cite{bw0409}.  The ${\cal O}(p^2)$
terms are potential higher-derivative terms.

\subsection{$\bf\nf=4$}

The next case is $\nf=4$.  This is the first case where we have 
a singular superpotential (\ref{spweff}).  Since we need a total 
of six external massless legs, we can only have one diagram (plus
its various corssings) with an internal chiral vertex.  This is
the single diagram shown in figure 2b.  There are also a number of
purely anti-chiral diagrams which could contribute.  We will show,
quite generally, that these diagrams vanish in the $\ve\to0$ limit,
leaving only the diagram in figure 2b.

\FIGURE{
\epsfig{file=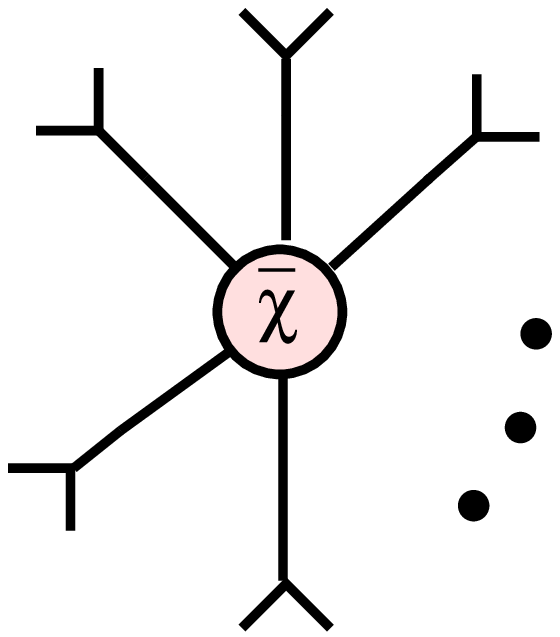,width=7em}      
\caption{The sum of all purely anti-chiral diagrams vanishes
for $\nf>3$.}}

We now show that the sum of all purely anti-chiral diagrams, 
represented in figure 3, vanishes for $\nf>3$.  All but one of 
the legs of the $\bar\chi$ subdiagram has a $D^2$ by the Feynamn 
rules.  Rewriting the overall Grassmann integration for $\bar\chi$ 
as $\int d^4\th=\int d^2\thb D^2$ gives the remaining leg a $D^2$.  
These $D^2$'s combine with the $\Db^2$'s from each anti-chiral 
propagator connecting $\bar\chi$ to the external vertices to give 
a factor of $p^2$ which cancels the $p^2$ in the denominators of 
those propagators.  Thus all the potential IR poles cancel, leaving 
no $D$'s or $\Db$'s to act on the massless external background fields 
on the external legs.  

So, setting the momenta to zero gives a finite result.  But, in the
$\ve\to0$ limit, an $SU(2\nf-2)$ subgroup of the global flavor
symmetry is restored.  So, the coefficient of the leading power of
$\ve$ will be $SU(2\nf-2)$-invariant.  Thus the leading term in
the $p^2\to0$ and $\ve\to0$ limit of the sum of all diagrams
of the form shown in figure 3 will be proportional to
\be\label{amp3b}
\e^{u_1v_1\cdots u_{\nf-1}v_{\nf-1}}
\dMb_{1u_1}\dMb_{2v_1}\cdots \dMb_{1u_{\nf-1}}\dMb_{2v_{\nf-1}},
\ee
since the completely antisymmetric tensor is the only
$\SU(2\nf-2)$-invariant way of tying together the flavor indices
of the massless external $\dMb$ fields.  But the expression in
(\ref{amp3b}) vanishes since the product of the $\dMb_{1u_i}$'s
and that of the $\dMb_{2v_i}$'s are symmetric on their $u_i$ and $v_i$
indices, respectively.

But this is only the leading term in an expansion around $p^2=0$.  
Higher powers of $p^2$ can be brought to act on the external legs,
giving derivatives of the external fields in the combinations 
$\del^2(\dMb_{1u_i}\dMb_{2v_i})$.  The higher powers of $p^2$ 
come from the Taylor expansion of the $(1+|\ve^\a|^2p^2)^{-1}$
denominators of the propagators (\ref{props}).  Thus each
factor of $p^2$ comes with a factor of $|\ve|^{2\a}$.  The flavor
symmetry of the leading term in the $\ve$-expansion of the amplitude
ensures that the external $u_i$ and $v_i$ indices are completely 
antisymmetrized.  This still enforces the vanishing of the amplitudes 
as long as there are at least two factors of $(\dMb_{1u_i}\dMb_{2v_i})$
without derivatives acting on them.  Thus, the first non-vanishing
term will have a factor of $p^2$ acting on $\nf-2$ pairs of external
legs.  

Now consider any purely anti-chiral internal sub-diagram $\bar\chi$.  
Each anti-chiral vertex has a $D^2$ acting on all but one of its legs 
as well as an $\int d^4\th_i$.  Likewise each internal anti-chiral 
propagator has a $\Db^2$ as well as a $\d^4(\th_i-\th_j)$.  The delta 
functions and Grassmann integrations leave just a single overall 
$\int d^4\th$.  The $\Db^2$'s and $D^2$'s pair up so there is a 
$\Db^2D^2=p^2$ in the numerator of each internal chiral propagator, 
and a $D^2$ acting on all but one of the external legs.  This $p^2$ 
cancels the $p^2$ in the denominator of the anti-chiral propagator in 
(\ref{props}), leaving the IR-finite factor proportional to 
$\veb^\a(\mb)^{-\b}(\lb)^{-1}$.

If the purely anti-chiral sub-diagram has $P$ internal propagators,
$E$ external legs, and $V_{\ell,0}$ $\ell$-legged vertices, 
this implies that the whole sub-diagram gives an effective vertex 
proportional to 
\be\label{3.14}
\lb^{(-P   + \sum_\ell             V_{\ell,0})}
\veb^{( P\a - \sum_\ell \g_{\ell,0} V_{\ell,0})}
\mb^{(-P\b - \sum_\ell \k_0        V_{\ell,0})}
= \lb \veb^{-E+(n+1)/n} \mb^{1/n},
\ee
plus terms vanishing as $p^2\to0$.  On the right side
we have substituted the values of $\a$, $\b$, $\g_{\ell,0}$, 
and $\k_0$ from (\ref{abeq}), (\ref{gkeq}), and used the
identities
\be\label{treerules}
P+1 = \sum_{\ell,m} V_{\ell,m}, \qquad
2P+E = \sum_{\ell,m} \ell V_{\ell,m},
\ee
where $V_{\ell,m}$ is the number of vertices with a total of 
$\ell$ legs of which $m$ are massless external legs.  They
follow from the topology of connected tree diagrams.
(We have set the number $m$ of massless external legs to zero 
because our sub-diagram is internal, so only connects to massive 
propagators.)

Now we can compute the dependence on $\ve$ of the purely 
anti-chiral amplitude in figure 3 with $N$ factors of $p^2$;
it will have an overall factor of $\ve$ to the power of
\be
2\a N + \left({n+1\over n}-E\right) + (\nf-1)\a - (\nf-1)\g_{3,2}
= (2 N - \nf+1){n-1\over n},
\ee
where the first term is from the $N$ factors of $p^2$, the second from 
the $\bar\chi$ internal diagram (\ref{3.14}) with $E=(\nf-1)$ legs, the 
third from the $(\nf-1)$ anti-chiral propagators attaching $\bar\chi$ to 
the external 3-vertices and the fourth from the $(\nf-1)$ 3-vertices 
themselves each with 2 massless legs.  We have used the values of $\a$ 
and $\g_{3,2}$ from (\ref{abeq}) and (\ref{gkeq}) on the right-hand side.  
Thus the power of $\ve$ is non-negative when $N\ge (\nf-1)/2$.  The 
minimum value of $N=\nf-2$ needed for the amplitude not to vanish by 
antisymmetry is greater than $(\nf-1)/2$ for $\nf>3$.  Thus, for $\nf>3$ 
the sum of all the diagrams of the form shown in figure 3 vanish as 
$\ve\to0$.  We evaluated the special $\nf=3$ case above and saw
explicitly that it does not vanish.

\FIGURE{
\epsfig{file=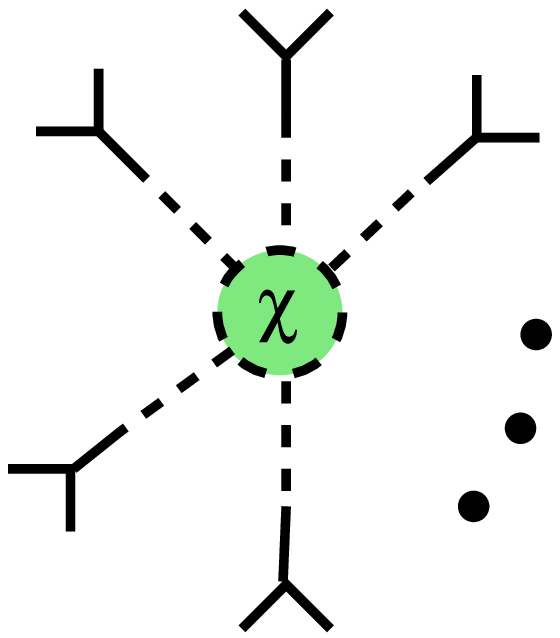,width=7em}      
\caption{The sum of all diagrams with purely chiral internal
vertices.}}

It remains to evaluate the single diagram in figure 2b.  It is a
special case of the class of diagrams shown in figure 4:  purely-chiral
internal diagrams with anti-chiral external 3-vertices.  It is
easy to evaluate the overall structure of these amplitudes.

The Feynman rules imply that there 
is a $\Db^2$ acting on all but one of the $\nf-1$ legs of the 
$\chi$ internal sub-diagram.  Rewriting the overall Grassmann 
integration for $\chi$ as $\int d^4\th=\int d^2\th \Db^2$ gives 
the remaining leg a $\Db^2$.  Thus each mixed propagator connecting 
the $\chi$ sub-diagram to the external anti-chiral 3-vertices
will have a $\Db^2$ acting on it.  Unlike the purely anti-chiral
propagator, the mixed propagator (\ref{props}) has neither an IR 
pole nor any $D^2$'s in the numerator.  Thus each $\Db^2$ will act 
on a pair of external massless legs.  To leading order in the 
$\dMb$'s, $\Db^2\dMb = 0$ by equation of motion, so we can replace 
$\Db^2(\dMb^2)=(\Db\dMb)^2$.  Thus, the massless external background 
fields must appear as
\be\label{amp3a}
\int d^4xd^2\th \, (\Db\dMb_{1u_1} \cdot \Db\dMb_{2v_1}) \cdots 
(\Db\dMb_{1u_{\nf-1}} \cdot \Db\dMb_{2v_{\nf-1}}).
\ee
As before, the leading term in the $\ve\to0$ limit must be
invariant under the $\SU(2\nf-2)$ subgroup of the flavor
symmetry that is not broken by the vacuum, and so the $u_i$ and
$v_i$ indices must be contracted with the totally antisymmetric
tensor $\e^{u_1v_1\cdots u_{\nf-1}v_{\nf-1}}$.

It is easy to compute the dependence of this amplitude on
$\l$, $\m$ and $\ve$.  With $E=\nf-1$ external legs, we get from 
the $\chi$ internal sub-diagram a factor, as in (\ref{3.14}),
\be
\l \ve^{1-\nf+(n+1)/n} \m^{1/n};
\ee
while the $\nf-1$ anti-chiral 3-point vertices with 2 massless 
legs contribute a factor, (\ref{3.8}) and (\ref{gkeq}),
\be
(\lb \veb^{-2+(n+1)/n} \mb^{-1+1/n})^{\nf-1};
\ee
and the $\nf-1$ mixed propagators at $p^2=0$ give the
factor, (\ref{props}),
\be
|\l^{-1}\ve^{2-(n+1)/n} \m^{-1/n}|^{2(\nf-1)}.
\ee
Combining all these factors with (\ref{amp3a}), and recalling that
$n=\nf-2$, gives 
\be\label{3.27}
\int d^4xd^2\th \, \l^{2-\nf}\mb^{1-\nf}\m^{-1}
\e^{u_1v_1\cdots u_{\nf-1}v_{\nf-1}}
(\Db\dMb_{1u_1} \cdot \Db\dMb_{2v_1})
\cdots (\Db\dMb_{1u_{\nf-1}} \cdot \Db\dMb_{2v_{\nf-1}}),
\ee
which is $\ve$-independent.  This expression, up to a numerical factor, 
coincides with (\ref{3.4}): the $\SU(2)$ superQCD higher-derivative
F-terms of \cite{bw0409}. 

Since this was the only diagram contributing in the $\nf=4$ case,
and since there is only a single diagram in that case, there can be 
no cancellation of its coefficient.  This shows that the $\nf=4$ 
singular superpotential indeed reproduces the corresponding 
higher-derivative global F-term in perturbation theory. 
With some more work, this argument could be turned into a 
calculation of the value of the coefficient of the higher-derivative
term.  But since the normalization of the higher-derivative F-terms 
was not determined in \cite{bw0409}, we are content to have simply 
shown that the coefficient is non-zero.

\subsection{$\bf\nf\ge5$}

\FIGURE{
\epsfig{file=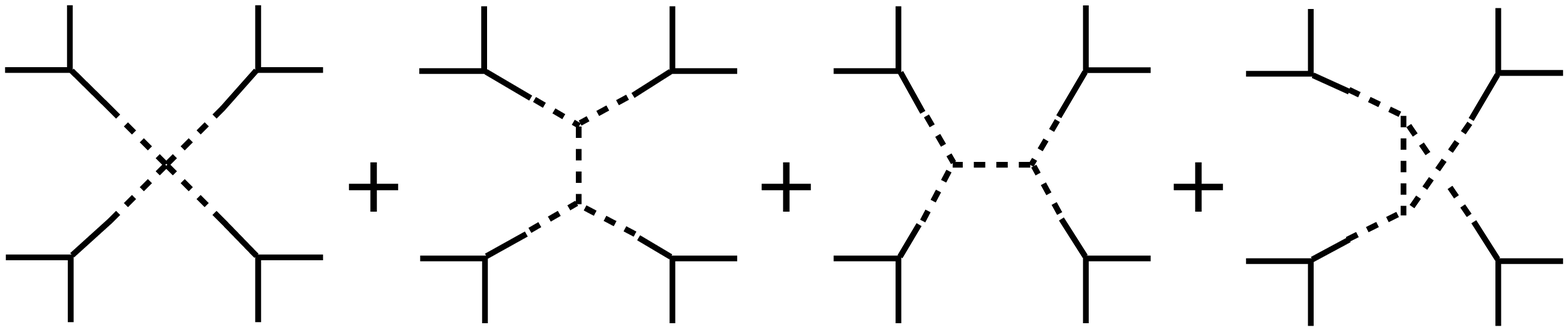,width=17em}     
\caption{Diagrams for $\nf=5$ flavors.}}

As we go higher in the number of flavors, however, the number of
diagrams contributing to each amplitude increases.  For instance, 
just among the class of internally purely-chiral diagrams illustrated
in figure 4, there are four Feynman diagrams in the case of 
$\nf=5$ flavors.   As sketched in figure 5, we have one diagram with 
a single internal vertex, and three different combinations of a diagram 
with two internal vertices.  Although we have shown above that the 
leading contribution of the sum of these diagrams has the right 
structure to reproduce the predicted higher-derivative F-term, 
since now multiple diagrams contribute, we must show in addition 
that no cancellations occur that could set the coefficient of the 
higher-derivative term to zero.  This seems quite complicated, as
it depends on the signs and tensor structures of the vertices.
Some sort of symmetry argument is clearly wanted, but still
eludes us.

In addition, there are now also other classes of diagrams which
are neither purely anti-chiral (as in figure 3) or internally
purely chiral (as in figure 4).  It is not clear whether these
mixed diagrams will also contribute to higher-derivative
amplitudes of the form (\ref{3.27}) or not.

\section{Singular superpotentials in three dimensions} 

It is worth mentioning that the method we developed here to get
the moduli space of the theory from the singular superpotential 
(\ref{spweff}) is not unique to four dimensions. In fact, as we will 
show below, the method can be used to obtain the moduli space of 
three dimensional supersymmetric gauge theories (with four supercharges) 
from singular superpotentials, wherever one is allowed to write 
such singular superpotenials.  See, for example, \cite{ahiss9703,
dho9703} for discussions of $N=2$ supersymmetric gauge theories in
three dimensions.

Consider an $N=2$ $\SU(2)$ supersymmetric gauge theory in three 
dimensions with $2\nf$ light flavors $Q^i_a$, transforming in the 
fundamental representation where $i=1,\cdots, 2\nf$ and $a=1,2$. 
Classically, the moduli space of the theory has a Coulomb branch 
as well as a Higgs branch for $\nf\neq 0$.  The Coulomb branch is 
parameterized by the vacuum expectation values of $U=e^\Phi$ where 
$\Phi$ is a chiral superfield.  The scalar component of $\Phi$ is 
$\phi +i\sigma$, where $\phi \in R /\Z_2$ is the scalar in the 
vector multiplet of the unbroken $\U(1)$ and $\sigma \sim \sigma + 
2\pi r$ is the scalar dual to the gauge field.  The Higgs branch 
is parameterized by the vacuum expectation values of $V^{ij}= \e^{ab}
Q^i_a Q^j_b$.  For $\nf=1$, $V^{ij}$ is unconstrained while for 
$\nf\geq 2$, $V^{ij}$ is subject to rank $(M)\leq 2$, or equivalently
\be\label{4.11}
\e_{i_1\ldots i_{2N_f}} V^{i_1 i_2} V^{i_3 i_4} = 0,
\ee
just as in the four-dimensional case.  

The quantum global symmetry of the theory is $\SU(2\nf)\times \U(1)_A 
\times\U(1)_R$ under which the fields parametrizing the Coulomb and 
the Higgs branch transform as
\be\label{b}
\begin{array}{c@{\hspace{.1in}\extracolsep{.1in}}ccccl}
               & \SU(2\nf ) & {\U(1)}_A  & {\U(1)}_R & \\[1ex]  
U              &  \bf 1    & -2\nf     & 2(1-\nf) & \\[1ex]
V^{ij}     	   &  \wedge^2({\bf 2\nf})  & 2 & 0 &. \\[1ex]
\end{array}
\ee
For $\nf\geq 2$, the quantum Higgs branch is the same as the classical 
Higgs branch, \ie\ it is described by (\ref{4.11}). We will be 
interested in the Higgs branch of the moduli space only for $\nf>2$ 
where the global symmetry of the theory requires one to consider the 
singular superpotential \cite{ahiss9703}
\be\label{4.12}
W =(1-\nf)( U\ \Pf V )^{1/(\nf-1)}. 
\ee    

Although this superpotential is singular, it describes the moduli
space perfectly for points away from the origin.  (There are additional
light degrees of freedom at the origin, which are not captured in
(\ref{4.12}).)  To show this, we have to first deform 
(\ref{4.12}), then send the deformation parameters to zero at the end. 
In close analogy to what we did in four dimensions in section 2, we 
deform $W$:
\be\label{4.13}
W\to W^{\z, \eta} = W + \z U + {1\over 2}\eta_{ij}V^{ij},  
\ee  
where $\z$ and $\eta_{ij}$ are some invertible parameters. 
The equations of motion for $U$ and $V^{kl}$ yield
\bea\label{4.14}
\z&=&\left(U^{2-\nf} \Pf V\right)^{1/(\nf-1)},
\nonumber\\
V^{kl}&=&-\left(U \ \Pf V\right)^{1/(\nf-1)} (\eta^{-1})^{kl}.
\eea  
Solving the first for $U$ and substituting the result into 
the second gives an equation for $V^{kl}$ which can be solved to obtain 
\be\label{4.17}
V^{kl}=-\left(\z\ \Pf\eta\right)^{1/2} (\eta^{-1})^{kl}.
\ee
Multiplying the above equation by itself and contracting the result
with $\e_{i_1\ldots i_{2\nf}}$, we arrive at 
\be\label{4.18}
\e_{i_1\ldots i_{2\nf}} V^{i_1 i_2} V^{i_3 i_4} =
\e_{i_1\ldots i_{2\nf}}\z\ \Pf\eta\ (\eta^{-1})^{i_1 i_2}\,
(\eta^{-1})^{i_3 i_4}.
\ee
The right hand side of (\ref{4.18}) is a polynomial of order 
$\nf-2 > 0$ for $\eta_{ij}$ and of order one for $\z$.  Therefore, 
independent of how we send  $\e_{ij}$ and $\z$ to zero, the right 
hand side of (\ref{4.18}) will vanish and we obtain 
\be\label{4.19}
\e_{i_1\ldots i_{2\nf}} V^{i_1 i_2} V^{i_3 i_4} = 0,
\ee
which is exactly (\ref{4.11}), the constraint equation describing 
the moduli space. 

This example gives some evidence that singular superpotentials can 
also perfectly-well describe the moduli space in supersymmetric gauge 
theories in three dimensions with four supercharges.  A similar
argument should also work to describe the moduli space for various 
$(2,2)$ supersymmetric gauge theories in two dimensions.  However, 
unlike the situation in four dimensions, there is no range of
flavors in these lower-dimensional theories where the theory is
IR free.  This makes a rigorous justification for the existence
of the effective superpotentials of these theories harder to
come by.  In certain cases, like the example discussed above,
the lower-dimensional theory can be obtained by compactification
of a four-dimensional theory on a circle.

\section*{Acknowledgments}
It is a pleasure to thank C. Beasely, M. Douglas, S. Hellerman, 
N. Seiberg, M. Strassler, P. Svr\v cek and E. Witten for helpful 
comments and discussions, and to thank the School of Natural Sciences
at the Insitite for Advanced Study, where much of this work was done,
for its hospitality and support.  
PCA and ME are supported in part by DOE grant DOE-FG02-84ER-40153.  
PCA was also supported by an IBM Einstein Endowed Fellowship.


\begin{thebibliography}{99}

\bibitem{bw0409}
C. Beasley and E. Witten,
{\sl New instanton effects in supersymmetric QCD,}
\jhep{0501}{2005}{056},
[\hepth{0409149}].

\bibitem{is9509}
K.A. Intriligator and N. Seiberg,
{\sl Lectures on supersymmetric gauge theories and electric-magnetic duality},
\npps{45BC}{1996}{1},
[\hepth{9509066}].

\bibitem{p9702}
M.E. Peskin, 
{\sl Duality in supersymmetric Yang-Mills theory},
[\hepth{9702094}].

\bibitem{s9402}
N. Seiberg,
{\sl Exact results on the space of vacua of four-dimensional SUSY gauge theories},
\prd{49}{1994}{6857}
[\hepth{9402044}].

\bibitem{ip9505}
K.A. Intriligator and P. Pouliot,
{\sl Exact superpotentials, quantum vacua and duality in supersymmetric 
Sp($N_c$) gauge theories},
\plb{353}{1995}{471},
[\hepth{9505006}].

\bibitem{s9411}
N. Seiberg,
{\sl Electric-magnetic duality in supersymmetric nonabelian gauge theories},
\npb{435}{1995}{129}
[\hepth{9411149}].

\bibitem{cdsw0211}
F. Cachazo, M.R. Douglas, N. Seiberg and E. Witten,
{\sl Chiral rings and anomalies in supersymmetric gauge theory},
\jhep{0212}{2002}{071}
[\hepth{0211170}].

\bibitem{s0212}
N. Seiberg,
{\sl Adding fundamental matter to `Chiral rings and anomalies in supersymmetric
gauge theory'},
\jhep{0301}{2003}{061},  
[\hepth{0212225}].

\bibitem{k84}
K. Konishi,
{\sl Anomalous supersymmetry transformation of some composite operators in SQCD},
\plb{135}{1984}{439}.

\bibitem{ks85}
K. Konishi and K.I. Shizuya, 
{\sl Functional integral approach to chiral anomalies in supersymmetric gauge theories},
{\sl Nuovo Cim.}\ {\bf A 90} (1985) 111.

\bibitem{vy82}
G. Veneziano and S. Yankielowicz,
{\sl An effective lagrangian for the pure N=1 supersymmetric Yang-Mills theory},
\plb{113}{1982}{231}.

\bibitem{ae}
P. C. Argyres and M. Edalati, in preparation.

\bibitem{w82}
E. Witten, 
{\sl An SU(2) anomaly}, 
\plb{117}{1982}{324}.

\bibitem{ads84}
I. Affleck, M. Dine and N. Seiberg,
{\sl Dynamical supersymmetry breaking in supersymmetric QCD},
\npb{241}{1984}{493}.

\bibitem{c0305}
S. Corley,
{\sl Notes on anomalies, baryons, and Seiberg duality,}
[\hepth{0305096}].

\bibitem{w0302}
E. Witten,
{\sl Chiral ring of Sp(N) and SO(N) supersymmetric gauge theory in four
dimensions,}
[\hepth{0302194}].

\bibitem{s0311}
P. Svr\v cek,
{\sl On non-perturbative exactness of Konishi anomaly and the 
Dijkgraaf-Vafa conjecture},
\jhep{0410}{2004}{028}, [\hepth{0311238}].

\bibitem{ggrs0108}
S.J. Gates, M.T. Grisaru, M. Rocek and W. Siegel,
{\sl Superspace, or one thousand and one lessons in supersymmetry},
Benjamin/Cummings 1983,   
[\hepth{0108200}].

\bibitem{ahiss9703}
O. Aharony, A. Hanany, K.A. Intriligator, N. Seiberg and M.J. Strassler,
{\sl Aspects of N = 2 supersymmetric gauge theories in three dimensions},
\npb{499}{1997}{67}, 
[\hepth{9703110}].

\bibitem{dho9703}
J. de Boer, K. Hori and Y. Oz,
{\sl Dynamics of N = 2 supersymmetric gauge theories in three dimensions},
\npb{500}{1997}{163},  
[\hepth{9703100}].

\end{thebibliography}
\end{document}